\documentclass[10pt]{bmc_article}

\usepackage{cite} 
\usepackage{url}  
\usepackage{ifthen}  
\usepackage{multicol}   
\usepackage[utf8]{inputenc} 
\usepackage{graphicx}
\usepackage{amsmath,amsfonts} 
\newboolean{publ}

\newcommand{\be}{\begin{eqnarray}}
\newcommand{\ee}{\end{eqnarray}}
\newcommand{\ba}{\begin{array}}
\newcommand{\ea}{\end{array}}
\newcommand{\bv}{\begin{verbatim}}
\newcommand{\ev}{\end{verbatim}}
\newcommand{\bt}{\begin{tabular}}
\newcommand{\et}{\end{tabular}}
\newcommand{\btab}{\begin{table}}
\newcommand{\etab}{\end{table}}
\newcommand{\bfig}{\begin{figure}}
\newcommand{\efig}{\end{figure}}
\newcommand{\bc}{\begin{center}}
\newcommand{\ec}{\end{center}}
\newcommand{\bit}{\begin{itemize}}
\newcommand{\eit}{\end{itemize}}
\newcommand{\nn}{\nonumber}

\newcommand{\tw}{\textwidth}
\newcommand{\ig}[1]{\includegraphics[width=#1\tw]}

\newcommand{\arr}{$\rightarrow$ }

\newenvironment{bmcformat}{\baselineskip20pt\sloppy\setboolean{publ}{false}}{\baselineskip20pt\sloppy}

\begin{document}
\begin{bmcformat}


\title{Phylogenetic tree reconstruction from genome-scale metabolic models}
 

\author{D.~Gamermann\correspondingauthor$^{1,2}$%
         \email{Daniel Gamermann\correspondingauthor - daniel.gamermann@ucv.es}
       \and 
        A.~Montagud$^2$
       \and
        J.~A.~Conejero$^2$
       \and
        P.~F.~de C\'ordoba$^2$
       and
        J.~F.~Urchiegu\'ia$^2$
      }


\address{%
    \iid(1)C\'atedra Energesis de Tecnolog\'ia Interdisciplinar, Universidad Cat\'olica de Valencia San Vicente M\'artir, \\ Guillem de Castro 94, E-46003, Valencia, Spain.\\
    \iid(2)Instituto Universitario de Matem\'atica Pura y Aplicada, Universidad Polit\'ecnica de Valencia, \\  Camino de Vera 14, 46022 Valencia, Spain.
}%

\maketitle


\begin{abstract}
A wide range of applications and research has been done with genome-scale metabolic models. In this work we describe a methodology for comparing metabolic networks constructed from genome-scale metabolic models and how to apply this comparison in order to infer evolutionary distances between different organisms. Our methodology allows a quantification of the metabolic differences between different species from a broad range of families and even kingdoms. This quantification is then applied in order to reconstruct phylogenetic trees for sets of various organisms. 
\end{abstract}

\ifthenelse{\boolean{publ}}{\begin{multicols}{2}}{}


\section{Introduction}

Metabolic models at the genome-scale are one of the pre-requisites for obtaining insight into the operation and regulation of metabolism as a whole \cite{arn1,arn2,arn3,arn4}. Uses of metabolic models embrace all aspects of biotechnology: from food \cite{arn5} to pharmaceutical \cite{arn6} and biofuels \cite{arn7,arn8,arn9}. Genome-scale metabolic network reconstruction is, in essence, a systematic assembly and organization of all reactions which build up the metabolism of a given organism. It usually starts with genome sequences in order to identify reactions and network topology. This methodology also offers an opportunity to systematically analyse {\it omics} datasets in the context of cellular metabolic phenotypes. 

Reconstructions have now been built for a wide variety of organisms, and have been used toward five major ends \cite{arn10}: contextualization of high-throughput data \cite{arn4,arn8,arn9,arn11}, guidance of metabolic engineering \cite{arn12}, directing hypothesis-driven discovery \cite{arn15}, interrogation of multi-species relationships \cite{arn16}, and network property discovery \cite{arn13}. 

Nowadays, phylogenies are becoming increasingly popular, being used in almost every branch of biology \cite{arn14}. Beyond representing the relationships among species in the tree of life, phylogenies are used to describe relationships between paralogues in a gene family \cite{arn15}, histories of populations \cite{arn16}, the evolutionary and epidemiological dynamics of pathogens \cite{arn17, arn18}, the genealogical relationship of somatic cells during differentiation and cancer development \cite{arn19} and even the evolution of language \cite{arn20}. More recently, molecular phylogenetics has become an indispensible tool for genome comparisons \cite{arn21,arn22,arn23}.

A phylogeny is a tree containing vertex that are connected by branches. Each branch represents the persistence of a genetic lineage through time, and each vertex represents the birth of a new lineage. If the tree represents the relationships among a group of species, then the vertex represent speciation events. Phylogenetic trees are not directly observed and are instead inferred from sequence or other data. Phylogeny reconstruction methods are either distance-based or character-based. In distance matrix methods, the distance between every pair of sequences is calculated, and the resulting distance matrix is used for tree reconstruction. For a very instructive review, please refer to \cite{arn14}.

This work is organized as follows. In the next section we briefly explain the genome-scale models we work with and how we define a parameter for comparing two models. Following that section, we explain how we recover the phylogenetic tree from the comparison matrix obtained for many metabolic models. This will be done taking into account the minimum spanning tree of a non-directed connected weighted network associated to these metabolic models.
In the section after that, we present the results and a brief study of the sensibility of the comparison parameter, and in the last section we present a brief summary and overview.


\section{Comparison between metabolic models}

In a recent paper \cite{raymari}, it has been presented a method for automatically generate genome-scale metabolic models from data contained in the KEGG \cite{kegg} database. The method consists in searching the database for genes and pathways present in an organism and download the corresponding set of chemical reactions. The algorithm filters isosenzymes, or other repeated reactions, and may add missing reactions to a given pathway using a probabilistic criterion based on the comparison of the organism's pathway with the same pathway in other organisms. In this work we are using data obtained from this platform, but the method described can, in principle, be used with any set of metabolic models given that the compounds names in the models follow the same standard (the same compound has the same name in all models).

The first step in our work is to construct for every metabolic model $A$ a non-directed connected network $N_A=(V_A,E_A)$ from the information contained in it.
Here,  $V_A$ stands for the set of vertex  (nodes) of $A$ and $E_A$ for its set of edges. 
A metabolic model comprises a set of chemical reactions. Each chemical reaction associates a set of substrates with a set of products. For constructing the network, first we define the set of vertex $V_A$ as the set of compounds in $A$ (metabolites present in the model), 
assigning a vertex to each metabolite. The chemical reactions in the model will define the edges (links) of the network. If two metabolites appear as substrate and as a product, respectively, in a chemical reaction, an edge connecting the correspondent vertex is added to the network. A typical metabolic model of a prokaryote, with around 1000 metabolites and the same number of chemical reactions, becomes through this process a non-directed connected network with 1000 vertex and around 3000 links.

The problem at hand now is to elaborate a method for systematically compare and quantify the differences between two metabolic networks. For this purpose we define a parameter that scales between zero and infinity, zero meaning identical networks and infinity for networks that either share no node or no edge in common. The definition of this parameter is based on the identity of the nodes (the compounds), but not directly on the chemical reactions of the metabolic models, only indirectly through the edges of the network. 

Here we start with the metabolic networks of two organisms $A=(V_A,E_A)$ and $B=(V_B,E_B)$. The set of all metabolites in between the two organisms $A\cup B=(V_A\cup V_B,E_A\cup E_B)$ can be divided into a partition of three disjoint sets: the set of metabolites only present in $A$, the set of metabolites only present in $B$, and the set of metabolites common to both organisms:

\be
V_{A\cup B}&=& \underbrace{(V_A\setminus V_B)}_{\textrm{Only in A}}  \cup  \underbrace{(V_A\cap V_B)}_{\textrm{Common}} \cup \underbrace{(V_B\setminus V_A)}_{\textrm{Only in B}}
\ee
where $\setminus$ stands for the difference of sets. 
A representation of this situation is shown in Figure \ref{fig1}.
As it is represented there, each metabolite may have connections to metabolites within its set and connections to metabolites in the other sets.

Suppose that $V_A\cup V_B=\{v_1,\ldots,v_n\}$. Fix an arbitrary node $v_i$, $1\le i\le n$. We can consider its degree in $A\cup B$, i.e. the total number of connections of $v_i$ to the rest of metabolites of $V_A\cup V_B$, that we denote by $\deg(v_i)$. We can also consider the degree of $v_i$ when we restrict ourselves to the subnetwork generated by the vertex in $(V_A\setminus V_B)\cup\{v_i\}$, that we will call $\deg_{A\setminus B}(v_i)$. Similarly, we can also define $\deg_{A\cap B}(v_i)$, and $\deg_{B\setminus A}(v_i)$. With these degrees we can define, for each metabolite $v_i\in V_A\cup V_B$, the rate $p_{A\setminus B,i}$ of connections of $v_i$ to metabolites inside $A$ and not in $B$ respect to the total number of connections of $v_i$, that is: 
$$p_{A\setminus B,i}=\frac{\deg_{A\setminus B}(v_i)}{\deg(v_i)}.$$ Analogously, we can define $$p_{B\setminus A,i}=\frac{\deg_{B\setminus A}(v_i)}{\deg(v_i)} \quad \textrm{ and } \quad p_{A\cap B,i}=\frac{\deg_{A\cap B}(v_i)}{\deg(v_i)}.$$

The following weighted sum of the rates $p_{A\setminus B,i}$ provides a parameter of the differentiation of $A\cup B$ respect to $A$: 
$$
\alpha = \left(\frac{1}{|V_A\setminus V_B|}\sum_{v_j\in V_A\setminus V_B}\deg(v_j) \right)\sum_{v_i\in V_{A\setminus B}}\frac{p_{A\setminus B,i}}{\deg(v_i)}\nn
$$
On the one hand, the rates $p_{A\setminus B,i}$ are multiplied by the inverse of the total number of connections of $v_i$ in order to give more importance to the metabolites with fewer connections. The reason to do this is that metabolic networks of all organisms usually share their hubs (metabolites with many connections), so in order to establish differences and similitudes for different networks, one should focus on specific metabolites particular to only some organisms sharing common features.

On the other hand, the factor $\frac{1}{|V_A\setminus V_B|}\sum_{v_j\in V_A\setminus V_B}\deg(v_j)$  gives an average of the number of connections of the metabolites only present in $A$ respect to the whole network. This is done in order to rescale the size of the network.

Analogously, we can define $\beta$ and $\gamma$ from the metabolites in the other two sets.
$$
\beta = \left(\frac{1}{|V_B\setminus V_A|}\sum_{v_j\in V_B\setminus V_A}\deg(v_j) \right)\sum_{v_i\in V_{B\setminus A}}\frac{p_{B\setminus A,i}}{\deg(v_i)}\nn
$$
$$
\gamma =\left( \frac{1}{|V_A\cap V_B|}\sum_{v_j\in V_A\cap V_B}\deg(v_j) \right)\sum_{v_i\in V_{A\cap B}}\frac{p_{A\cap B,i}}{\deg(v_i)}\nn
$$

For illustrating the process, let's consider three organisms, the {\it Synechocystis} sp. PCC 6803 ( that we refer to as syn), {\it Synechococcus elongatus} PCC7942 (referred to as syf), and the {\it Escherichia coli} K-12 MG1655 (referred to as eco). In Table \ref{tab:orgsnet} one can see the number of metabolites and edges in the networks of these organisms, 
and in Table \ref{tab:sets} we show the number of elements in each one of the three sets of the partition in which we split the set of vertex of the network obtained from each pair of these three organisms.

Let's now focus on a few metabolites to see their contribution to the differentation parameters (i.e. to the parameters $\alpha$, $\beta$, and $\gamma$). For this, we chose Pyruvate (PYR), Glyoxylate (GXL), and 2-Dehydro-3-deoxy-6-phospho-D-gluconate (6PDG), which are respectively very, medium, and poorly connected metabolites present in these three organisms. In Table \ref{tab:metabs} we show the contribution of these metabolites to the parameters $\alpha$, $\beta$, and $\gamma$.

Finally, the comparison between the networks $A$ and $B$, namely $\zeta_{A,B}$,  is defined as:
\be
\zeta_{A,B}&=& \frac{\frac{|V_B|}{|V_A|}\alpha+\frac{|V_A|}{|V_B|}\beta}{2\gamma}\nn
\ee
The parameters $\alpha$ and $\beta$ are balanced since some organisms have much smaller metabolic networks than others.
If this is not corrected, it results in a disproportionate size between subnetworks generated by $V_{A\setminus B}$ and 
by $V_{B\setminus A}$. In order to weaken this difference, the parameter factors 
$\frac{|V_B|}{|V_A|}$ and $\frac{|V_A|}{|V_B|}$ are introduced.
For two identical networks, $\alpha$ and $\beta$ are zero, and so that $\zeta=0$. For two networks which have not a single metabolite in common we have $\gamma$=0 and so $\zeta=\infty$. 


\section{Reconstruction of the phylogenetic tree}
Given a set of $n$ organisms $\{A_1,A_2,\ldots,A_n\}$, we will see how to construct their phylogenetic tree taking into account the degrees of similarity between every pair of metabolic models.

Firstly, let $N=(V,E,w)$ be a non-directed connected complete weighted network, where $V=\{A_1,A_2,\ldots,A_n\}$ is the set of vertex that represent the metabolic models of the aforementioned organisms, $E$ is the set of edges
${(A_i,A_j),\, 1\le i,j \le n,\, i\ne j}$, and $w:E\rightarrow \mathbb{R}$ is a function that assigns to every edge $(A_i,A_j)$ the amount $w_{i,j}=\zeta_{A_i,A_j}$. Looking at the definition of $\zeta$, we observe that this network $N$ must be symmetric.
In particular, all the weights in our study are strictly positive.

Secondly, we will compute a minimum spanning tree of $N$, that is a tree which has $V$ as the set of vertex, and such that the sum of the weights associated to the edges of this tree is minimum. In these trees, every vertex $A_i\in V$ is connected with at least one of the other vertex of $V\setminus\{A_i\}$ by an edge that has minimum weight among all the edges incident to $A_i$.
The well-known Kruskal algorithm give us a procedure for finding these trees, see for instance \cite{gy}.
We just have to follow the trace of Kruskal algorithm in order to recover the phylogenetic tree of the organisms represented by the models $A_1,\ldots,A_n$. 

In order to compute the phylogenetic tree of the models $\{A_1,A_2,\ldots,A_n\}$, consider the minimum spanning tree of $N$, namely $T=(V,E',w|_{E'})$ where $E'\subset E$ and $w|_{E'}$ denotes the restriction of the function $w$  to the elements in $E'$. 
Let us take all the elements of $E'$ in decreasing order of weights, that is $E'=\{e'_1,e'_2,\ldots,e'_{n-1}\}$
with $w(e'_1)\ge w(e'_2) \ldots \ge w(e'_1)$. We are going to remove edges from $T$ following this order. 
Everytime an edge is removed, the number of connected components of the resulting graph is increased in one respect to the previous one. 
We can represent this division of connected componets by a binary tree. The phylogenetic tree is generated taking into account how we divide $T$.

There are two different situations depending on the size of the (new) connected components (if any of them consists on a single vertex or not).
Let us start with the edge with maximum weight in $T$, that we have denoted as $e'_1$. Suppose that $e'_1$ is adjacent to two vertex $A_{i_0}$ and $A_{j_0}$, with $1\le i_0,j_0 \le n$, $i\ne j$.
Then two situations can happen:
\begin{enumerate}
\item[(a)] One of these vertex, for instance $A_{i_0}$, is a leaf (vertex of degree 1), 
\item[(b)] No one of these two vertex is a leaf (each vertex is still connected with other vertex). This happens only if the former connected component has 3 or more vertex.
\end{enumerate}

We point out that our phylogenetic tree will have two types of vertex:
the leaves, that represent metabolic models, and the inner vertex, that represent that there are two branches and each one has more than one vertex.

We start our phylogenetic tree with a vertex $v_0$ that will be its root.
Then two vertex $v_1,v_2$ are hanged from $v_0$. Each one of these vertex represents one of the two connected components of the network $T\setminus\{e'_1\}$. Let us see what to do with $v_1$ and $v_2$ depending in which case we are.

\begin{itemize}
\item If we are in case (a), one of these two vertex, for instance $v_1$, represents the vertex $A_{i_0}$ and  $v_2$ the other one, $v_2$, represents the other connected component of $T$, which is generated subgraph of $T$ generated by the vertex of $V\setminus \{A_{i_0}\}$. 
\item If we are in case (b), one of the vertex, for instance $v_1$, represents the connected component of $T\setminus\{e'_1\}$ that contains $A_{i_0}$, and the other vertex, $v_2$, the connected component of $T\setminus\{e'_1\}$ that contains $A_{j_0}$. 
\end{itemize}

This procedure is repeated again with $v_1$ and $v_2$ and removing $e'_2$ from $T\setminus\{e'_1\}$. 
When we remove $e'_2$, then either the connected component that represents $v_1$ or $v_2$ is split into two smaller ones, and the vertex associated to this component plays again the role of $v_0$. 
This proccess is repeated until we remove all the edges.

Let us see how it works with two examples:
\begin{enumerate}
\item In Table \ref{tab1} we have the weights associated to a set of 10 organisms. The edge of maximum weight in the minimum spanning tree associated to the underlying network is the one that connects the mge with the lpl, with weight 0.123. 
We can see in Figure \ref{fig2} that two vertex are hanging from the root of the tree. The one on the left represents the \emph{mge}, the one on the right represents the subgraph associated to the rest of vertex, where the lpl can be found.
\item In the case of 33 organisms, when we remove from the minimum spanning tree the edge with maximum weight, we split this tree into 2 connected components: the one associated to the pair \emph{mge} and \emph{mpm}, 
and one the associated to the other vertex. 
\end{enumerate}

Finally, the vertex in the phylogenetic tree can keep more information concerning the aforementioned minimum spanning tree. 
Suppose that the height of our phlyogenic tree is $w(e'_1)$, that represents the maximum of the weights in the minimum spanning tree. i.e. the weight associated to $e'_1$. We place the root of our phylogenetic tree at height $y=w(e'_1)$.  Now, two vertex are hanged from the root. If one is associated to a single vertex for instance $v_1$ in case (a), then we place this vertex at height $y=0$. We remember that this vertex represents the metabolite  $A_{i_0}$. If not, for instance $v_2$ in case (a) and either $v_1$ or $v_2$ in case (b), each one of these vertex represents a connected component with more than one vertex in which the minimum spanning tree is split.
In order to know at which height we should put these vertex we have to continue removing edges from the former tree. After removing $e'_2$, one of these connected components, for instance the one represented by $v_2$, is split again into 2 smaller connected components. So we place the vertex $v_2$ at height $w(e'_2)$. We repeat this process recursevely until the initial tree is just reduced to isolated vertex.


\section{Results}

We have reconstructed two phylogenetic trees, one with 10 bacteria and another one with both, prokaryotes and eukaryotes. In Table \ref{tab1} we show the parameter $\zeta$ for the pairwise comparison of the 10 prokaryotes in the first tree. The data for the comparison of the 33 organisms in the second tree is given in a file in the supplementary materials. 

The organisms in each comparison are:

\bit
\item 10 organisms tree \arr {\it Mycoplasma genitalium} (mge), {\it Lactobacillus plantarum} WCFS1 (lpl), {\it Synechocystis} sp. PCC 6803 (syn), {\it Synechococcus elongatus} PCC7942 (syf), {\it Synechococcus elongatus} PCC6301 (syc), {\it Clostridium beijerinckii} (cbe), {\it Burkhoderia cenocepacia} J2315 (bcj), {\it Escherichia coli} K-12 MG1655 (eco), {\it Thermotoga maritima} (tma) and {\it Yersinia pestis} KIM10 (ypk).
\item 33 organisms tree \arr {\it Mycoplasma genitalium} (mge), {\it Mycoplasma pneumoniae} 309 (mpm), {\it Synechocystis} sp. PCC 6803 (syn), {\it Synechococcus elongatus} PCC7942 (syf), {\it Synechococcus elongatus} PCC6301 (syc), {\it Clostridium beijerinckii} (cbe), {\it Salmonella bongori} (sbg), {\it Escherichia coli} K-12 MG1655 (eco), {\it Aquifex aeolicus} (aae), {\it Yersinia pestis} KIM 10 (ypk), {\it Cyanobacterium} UCYN-A (cyu), {\it Thermosynechococcus elongatus} (tel), {\it Microcystis aeruginosa} (mar), {\it Cyanothece} sp. ATCC 51142 (cyt), {\it Cyanothece} sp. PCC 8801 (cyp), {\it Gloeobacter violaceus} (gvi), {\it Anabaena} sp. PCC7120 (ana), {\it Anabaena azollae} 0708 (naz), {\it Prochlorococcus marinus} SS120 (pma), {\it Trichodesmium erythraeum} (ter), {\it Acaryochloris marina} (amr), {\it Halophilic archaeon} (hah), {\it Polymorphum gilvum} (pgv), {\it Micavibrio aeruginosavorus} (mai), {\it Agrobacterium radiobacter} K84 (ara), {\it Clostridiales genomosp.} BVAB3 (clo), {\it Gamma proteobacterium} HdN1 (gpb), {\it Vibrio fischeri} ES114 (vfi), {\it Vibrio fischeri} MJ11 (vfm), {\it Haemophilus influenzae} F3031 (hif), {\it Coprinopsis cinerea} (cci), {\it Sus scrofa} (ssc) and {\it Leishmania braziliensis} (lbz).
\eit

In Figures \ref{fig2} and \ref{fig3} we present the two phylogenetic trees that we have constructed.

In the first tree the only organism displaced in relation to what is expected from standard methods of phylogenetic tree reconstruction is the tma. In both trees mge (and mpm in the second one) diverge from other organisms at the beginning of the tree. This happens because of their minimalistic genomes, with only a couple of hundred of metabolites in their metaboloms. As a result, when comparing with an organism without a reduced genome with around almost a thousand metabolites, several hundred metabolites will not have a correspondent one, increasing hugely the value of $\alpha$ in the calculation of the parameter $\zeta$, and therefore distancing these organisms from the rest. In any case, we used for the second study organisms from very different origins in the evolutionary history and we found that the method is able to separate bacteria, archea and eukaryotes. Different strains of the same species also appear closely related and sharing branches with organisms from the same family and order. 

We have also studied the sensibility of the parameter $\zeta$. For this we performed a monte-carlo analysis of $\zeta$. The procedure for this analysis is explained as follows. Given two organisms, one of them remains the wild type while, with the other, one builds an ensemble with $N_t$ elements, where each element is the result of $n_K$ knock-outs (removal of $n_K$ randomly selected reactions from the metabolic model) in the organism. Then the calculation of $\zeta$ is performed between the wild type organism and each organism in the knock-out ensemble. From this process one obtains an ensemble of $N_t$ values of $\zeta$ for the comparison (one from each version of the organism in the knock-out ensemble), from which one calculates its average and standard deviation. This standard deviation is treated as an indicator of the sensibility of the parameter (as a function the number of knock-outs).

We performed this sensibility analysis for four organisms (syn, syf, eco and mge) with ensembles of sizes $N_t=500$ for $n_K=$5, 10, 50 and 100. The results are in Tables \ref{tab2}-\ref{tab5}. These four organisms have been chosen in order to observe the sensibility in the comparison between very similar organisms (syn and syf), more distant ones (syn and eco) and very different ones (syn and mge).


\section{Conclusions}

In this work we have developed a methodology for comparing organisms based on their metabolic networks. This methodology has been successfully applied for the reconstruction of phylogenic trees for several organisms from a broad range of families and kingdoms. Resulting trees stand well their comparison with the so-called ``tree of life''. The great majority of the branches in the tree fit well their expected positions and their distance is in good correlation with evolutionary distances. The discrepancies found can be explained by particularities in these very few organisms not fitting the tree, such as tremendous genome reductions that caused reduced metabolisms.

Our methodology is innovative for it is not directly based on the structure and evolution of proteins or DNA, but on the metabolism and on the organisms' components and metabolic capabilities, allowing one to compare organisms very distant from the evolutionary point of view or organisms for which ortholog’s comparison is difficult. In order to accomplish this, we make use of the correlation between evolutionary distances and metabolic network likelihood and propose our methodology as a starting point to study it.

Metabolism information is retrieved as a subset of the whole genome information. We hereby show that metabolic network connectivity can be used to build phylogenetic trees that are in accordance with gene-directed trees. It can be argued whether the selected construction parameter ($\zeta$) is the optimal one for this purpose (or even if there is an optimal one), but it stands clear that this is an innovative application for metabolic models, their curation and cross-species evolutionary studies.

\bigskip

\section*{Author's contributions}
    D~.Gamermann calculated the comparison coefficients and together with A.~Conejero implemented the Kruskal algorithm and reconstructed the philogenetic trees. A.~Montagud analyzed the results and these three authors contributed to the writting of the manuscript. P~.F~. de C\'ordoba and J.~Urchuegu\'ia conceived and funded the study. All authors read and approved the manuscript.

\section*{Acknowledgements}
  \ifthenelse{\boolean{publ}}{\small}{}
  This work has been funded by the MICINN TIN2009-12359 project ArtBioCom from the Spanish Ministerio de Educaci\'on y Ciencia and FP7-ENERGY-2012-1-2STAGE (Project number 308518) CyanoFactory from the EU. 
 

\newpage
{\ifthenelse{\boolean{publ}}{\footnotesize}{\small}
 \bibliographystyle{bmc_article}  
  \bibliography{bmc_article} }     


\ifthenelse{\boolean{publ}}{\end{multicols}}{}



\section*{Figures}

\subsection{Figure 1 - conjs.pdf}\label{fig1}

\mbox{
\ig{0.5}{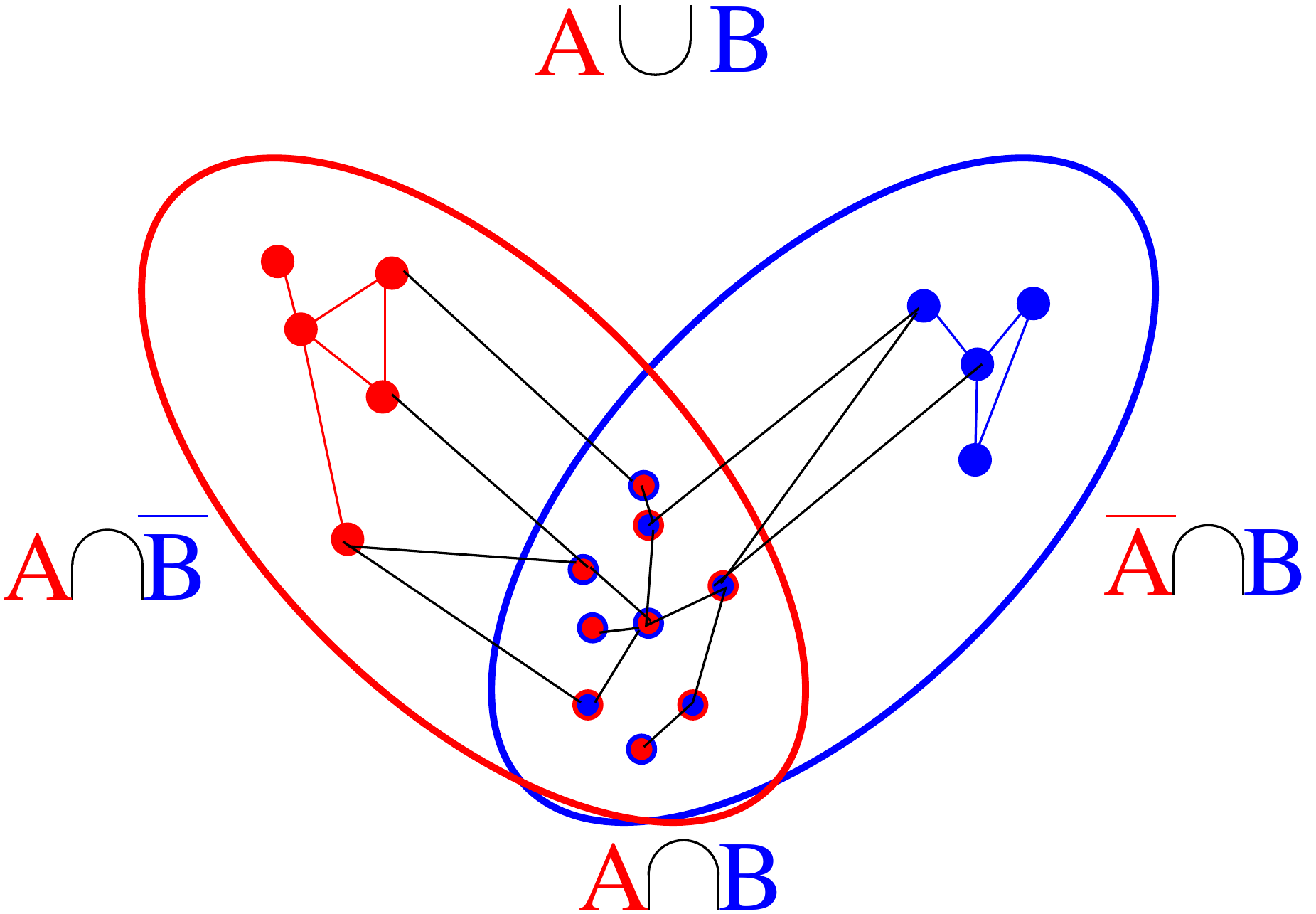}}

(Color on-line) Representation of the sets of metabolites between two organisms.

\subsection{Figure 2 - tree10.png}\label{fig2}

\mbox{
\ig{0.5}{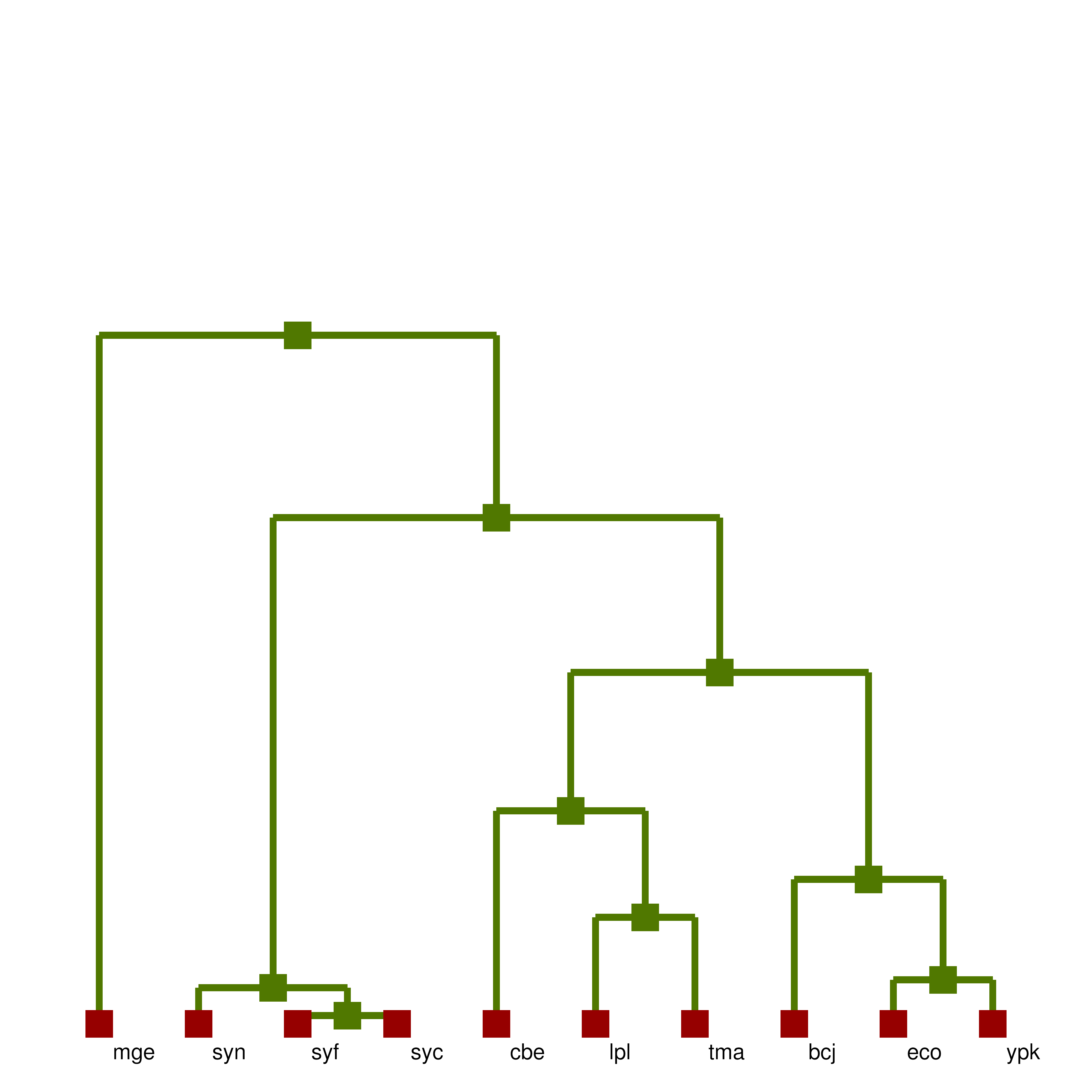}}

(Color on-line) Phylogenetic tree with 10 organisms.

\subsection{Figure 3 - tree30.png}\label{fig3}

\mbox{
\ig{1.0}{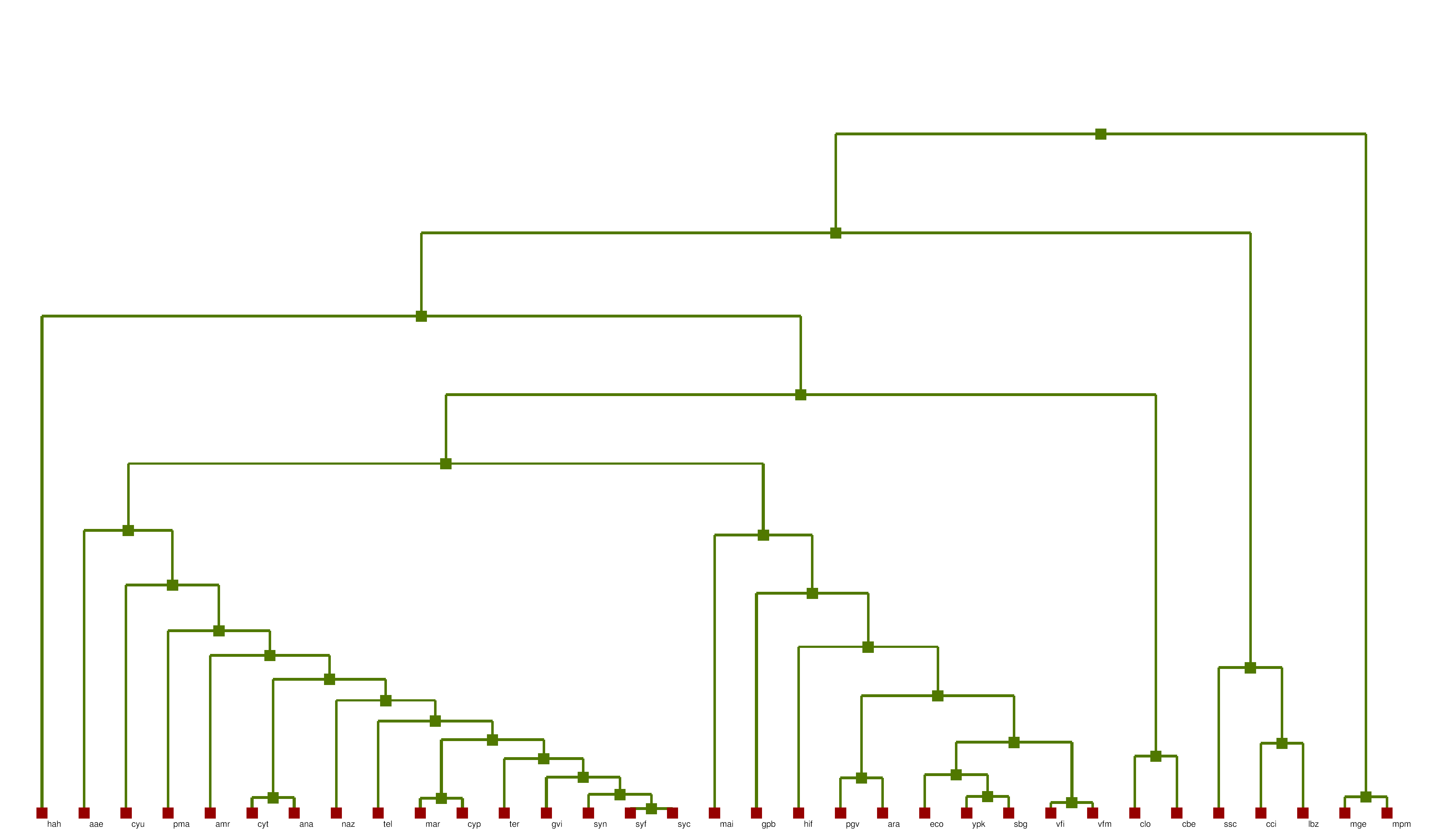}}

(Color on-line) Phylogenetic tree with 33 organisms.


\section*{Tables}

\subsection{Table 1 - Vertex and edges in the networks of syn, syf and eco.}\label{tab:orgsnet}

\mbox{
\bt{c|rr}
Organism & \# Vertex & \# Edges \\
\hline
syn & 1001 & 2891 \\
syf & 979 & 2810 \\
eco & 1227 & 3801 
\et
}

\subsection{Table 2 - Metabolites in the three sets of the partition when comparing three organisms.}\label{tab:sets}

\mbox{
\bt{c||l|l}
  & syf & eco \\
\hline
\hline
  & $|V_A \cap V_B|$ = 911 & $|V_A \cap V_B|$ = 778\\
 syn    & $|V_A \setminus V_B|$ = 90 & $|V_A \setminus V_B|$ = 223\\
     & $|V_B \setminus V_A|$ = 68 & $|V_B \setminus V_A|$ = 449\\
\hline
& - & $|V_A \cap V_B|$ = 775\\
 syf      & - & $|V_A \setminus V_B|$ = 204\\
     & - & $|V_B \setminus V_A|$ = 452\\
\et}

\subsection{Table 3 - Contribution of different metabolites to the differentiation parameter ($\zeta$) between two networks. The column $\delta_i$ shows the weight of the metabolite in the calculation of $p_{A\cap B,i}$ which is the inverse of the degree of the metabolite divided by the sum of the inverses of the degrees of all metabolites contributing to the parameter.}\label{tab:metabs}

\mbox{\bt{c|c|ccc}
Metabolite & Organisms in & $p_{A\cap B,i}$ & $\delta_i$ & Contribution (\%)\\
           & comparison   &                  & & \\
\hline
\hline
           & syn and syf  & 0.98 &  0.127 & 0.0064 \\
\cline{3-5}
PYR        & syn and eco  & 0.73 &  0.117 & 0.0044 \\
\cline{3-5}
           & syf and eco  & 0.75 &  0.113 & 0.0044 \\
\hline
           & syn and syf  & 0.86 &  0.454 & 0.020 \\
\cline{3-5}
GXL        & syn and eco  & 0.87 &  0.550 & 0.024 \\
\cline{3-5}
           & syf and eco  & 0.80 &  0.439 & 0.018 \\
\hline
           & syn and syf  & 1.00 &  3.176 & 0.16 \\
\cline{3-5}
6PDG   & syn and eco  & 0.80 &  1.762 & 0.072 \\
\cline{3-5}
           & syf and eco  & 0.80 &  1.757 & 0.072 \\
\hline
\et}

\subsection{Table 4 - Comparison matrix for ten organisms.}\label{tab1}

\mbox{
\bt{c|cccccccccc} 
 org       & mge     & lpl     & syn     & syf     & syc     & cbe     & bcj     & eco     & tma     & ypk     \\ 
\hline
\hline
mge & 0.0 & 0.123 & 0.1951 & 0.1753 & 0.1906 & 0.1438 & 0.1374 & 0.1384 & 0.1306 & 0.1384 \\ 
  \hline 
lpl & 0.123 & 0.0 & 0.1554 & 0.1543 & 0.1645 & 0.0719 & 0.1217 & 0.1227 & 0.0716 & 0.1154 \\ 
  \hline 
syn & 0.1951 & 0.1554 & 0.0 & 0.0195 & 0.0188 & 0.1286 & 0.1135 & 0.1137 & 0.1687 & 0.1248 \\ 
  \hline 
syf & 0.1753 & 0.1543 & 0.0195 & 0.0 & 0.0054 & 0.1262 & 0.117 & 0.1076 & 0.1677 & 0.1174 \\ 
  \hline 
syc & 0.1906 & 0.1645 & 0.0188 & 0.0054 & 0.0 & 0.1327 & 0.1149 & 0.1043 & 0.1657 & 0.1133 \\ 
  \hline 
cbe & 0.1438 & 0.0719 & 0.1286 & 0.1262 & 0.1327 & 0.0 & 0.1121 & 0.0933 & 0.0737 & 0.1071 \\ 
  \hline 
bcj & 0.1374 & 0.1217 & 0.1135 & 0.117 & 0.1149 & 0.1121 & 0.0 & 0.0678 & 0.1324 & 0.068 \\ 
  \hline 
eco & 0.1384 & 0.1227 & 0.1137 & 0.1076 & 0.1043 & 0.0933 & 0.0678 & 0.0 & 0.1101 & 0.0295 \\ 
  \hline 
tma & 0.1306 & 0.0716 & 0.1687 & 0.1677 & 0.1657 & 0.0737 & 0.1324 & 0.1101 & 0.0 & 0.1075 \\ 
  \hline 
ypk & 0.1384 & 0.1154 & 0.1248 & 0.1174 & 0.1133 & 0.1071 & 0.068 & 0.0295 & 0.1075 & 0.0 \\ 
  \hline 
\et
}

\subsection{Table 5 - Sensibility calculation for $N_t=500$ and $n_K=5$. Each element in the table is the average of the parameter $\zeta$ in an ensemble plus (minus) its standard deviation ($\bar{\zeta}\pm\sigma_\zeta$).}\label{tab2}

\mbox{
\bt{c|cccc}
org / org & syn & syf & eco & mge \\
\hline
syn & 0.0002 $\pm$ 0.0003  &   0.0184 $\pm$ 0.0005  &   0.0893 $\pm$ 0.0005  &   0.1600 $\pm$ 0.0014  \\
syf & 0.0184 $\pm$ 0.0004  &   0.0002 $\pm$ 0.0003  &   0.0857 $\pm$ 0.0006  &   0.1527 $\pm$ 0.0014  \\ 
eco & 0.0892 $\pm$ 0.0005  &   0.0856 $\pm$ 0.0005  &   0.0001 $\pm$ 0.0002  &   0.1278 $\pm$ 0.0009  \\ 
mge & 0.1597 $\pm$ 0.0025  &   0.1527 $\pm$ 0.0026  &   0.1283 $\pm$ 0.0015  &   0.0014 $\pm$ 0.0016  \\ 
\et}

\subsection{Table 6 - Sensibility calculation for $N_t=500$ and $n_K=10$. Each element in the table is the average of the parameter $\zeta$ in an ensemble plus (minus) its standard deviation ($\bar{\zeta}\pm\sigma_\zeta$).}\label{tab3}

\mbox{\bt{c|cccc}
org / org & syn & syf & eco & mge \\
\hline
syn & 0.0005 $\pm$ 0.0005  &   0.0186 $\pm$ 0.0006  &   0.0896 $\pm$ 0.0007  &   0.1604 $\pm$ 0.0018  \\ 
syf & 0.0187 $\pm$ 0.0006  &   0.0005 $\pm$ 0.0005  &   0.0860 $\pm$ 0.0007  &   0.1532 $\pm$ 0.0019  \\ 
eco & 0.0893 $\pm$ 0.0008  &   0.0857 $\pm$ 0.0007  &   0.0003 $\pm$ 0.0003  &   0.1281 $\pm$ 0.0011  \\ 
mge & 0.1602 $\pm$ 0.0035  &   0.1531 $\pm$ 0.0032  &   0.1288 $\pm$ 0.0023  &   0.0028 $\pm$ 0.0023  \\
\et}

\subsection{Table 7 - Sensibility calculation for $N_t=500$ and $n_K=50$. Each element in the table is the average of the parameter $\zeta$ in an ensemble plus (minus) its standard deviation ($\bar{\zeta}\pm\sigma_\zeta$).}\label{tab4}

\mbox{
\bt{c|cccc}
org / org & syn & syf & eco & mge \\
\hline
syn & 0.0028 $\pm$ 0.0011  &   0.0209 $\pm$ 0.0014  &   0.0915 $\pm$ 0.0017  &   0.1652 $\pm$ 0.0045  \\ 
syf & 0.0207 $\pm$ 0.0013  &   0.0029 $\pm$ 0.0011  &   0.0879 $\pm$ 0.0016  &   0.1575 $\pm$ 0.0044  \\ 
eco & 0.0903 $\pm$ 0.0017  &   0.0868 $\pm$ 0.0016  &   0.0016 $\pm$ 0.0007  &   0.1301 $\pm$ 0.0029  \\ 
mge & 0.1638 $\pm$ 0.0080  &   0.1577 $\pm$ 0.0077  &   0.1343 $\pm$ 0.0055  &   0.0170 $\pm$ 0.0053  \\
\et}

\subsection{Table 8 - Sensibility calculation for $N_t=500$ and $n_K=100$. Each element in the table is the average of the parameter $\zeta$ in an ensemble plus (minus) its standard deviation ($\bar{\zeta}\pm\sigma_\zeta$).}\label{tab5}

\mbox{\bt{c|cccc}
org / org & syn & syf & eco & mge \\
\hline
syn & 0.0058 $\pm$ 0.0016  &   0.0239 $\pm$ 0.0020  &   0.0942 $\pm$ 0.0024  &   0.1715 $\pm$ 0.0062  \\ 
syf & 0.0238 $\pm$ 0.0018  &   0.0061 $\pm$ 0.0017  &   0.0907 $\pm$ 0.0023  &   0.1630 $\pm$ 0.0066  \\ 
eco & 0.0919 $\pm$ 0.0024  &   0.0883 $\pm$ 0.0022  &   0.0033 $\pm$ 0.0011  &   0.1329 $\pm$ 0.0040  \\ 
mge & 0.1694 $\pm$ 0.0120  &   0.1648 $\pm$ 0.0131  &   0.1433 $\pm$ 0.0092  &   0.0460 $\pm$ 0.0076  \\
\et}


\section*{Additional Files}
  \subsection*{Additional file 1 - comps\_33orgs.dat}
    Table with the parameter $\zeta$ resulting from the comparison of 33 organisms.

\end{bmcformat}
\end{document}